\documentclass[a4paper,10pt,twoside]{cpc-hepnp}
\usepackage{CJK,upgreek,fancyhdr}
\usepackage{multicol}
\usepackage{graphicx}
\usepackage{booktabs}
\usepackage{amssymb,bm,mathrsfs,bbm,amscd}
\usepackage[tbtags]{amsmath}
\usepackage{lastpage}

\begin{document}
\begin{CJK*}{GB}{gbsn}


\footnotetext[0]{Received 9 May 2016}

\title{Cluster decay of the high-lying excited states in $^{14}$C}

\author{%
      Z.Y. Tian(ÌïÕýÑô)$^{1}$%
\quad Y.L. Ye (Ò¶ÑØÁÖ)$^{1;1)}$\email{yeyl@pku.edu.cn}%
\quad Z.H. Li(ÀîÖÇ»À)$^{1}$%
\quad C.J. Lin(Áֳмü)$^{2}$%
\quad Q.T. Li(ÀîÆæÌØ)$^{1}$\\%
\quad Y.C. Ge(¸ðÓä³É)$^{1}$%
\quad J.L. Lou(Â¥½¨Áá)$^{1}$%
\quad W. Jiang(½¯Î°)$^{1}$%
\quad J. Li(À)$^{1}$%
\quad Z.H. Yang(ÑîÔÙºê)$^{1}$\\%
\quad J. Feng(·ë¿¡)$^{1}$%
\quad P.J. Li(ÀîÅó½Ü)$^{1}$%
\quad J. Chen(³Â½à)$^{1}$%
\quad Q. Liu(ÁõÇ¿)$^{1}$%
\quad H.L. Zang(ê°ºêÁÁ)$^{1}$\\%
\quad B. Yang(Ñî±ë)$^{1}$%
\quad Y. Zhang(ÕÅÔÊ)$^{1}$%
\quad Z.Q. Chen(³Â־ǿ)$^{1}$%
\quad Y. Liu(ÁõÑó)$^{1}$%
\quad X.H. Sun(ËïÏþ»Û)$^{1}$\\%
\quad J. Ma(Âí¾º)$^{1}$%
\quad H.M. Jia(¼Ö»áÃ÷)$^{2}$%
\quad X.X. Xu(ÐìÐÂÐÇ)$^{2}$%
\quad L. Yang(ÑîÀÚ)$^{2}$%
\quad N.R. Ma(ÂíÄÏÈã)$^{2}$%
\quad L.J. Sun(ËïÁ¢½Ü)$^{2}$%
}
\maketitle

\address{%
$^1$ School of Physics and State Key Laboratory of Nuclear Physics and Technology, Peking University, Beijing, 100871, China\\
$^2$ China Institute of Atomic Energy, Beijing 102413, China\\
}

\begin{abstract}
A cluster-transfer experiment of $^9\rm{Be}(^9\rm{Be},^{14}\rm{C}\rightarrow\alpha+^{10}\rm{Be})\alpha$ at an incident energy of 45 MeV was carried out in order to investigate the molecular  structure in high-lying resonant states in $^{14}$C. This reaction is of extremely large $Q$-value, making it an excellent case to select the reaction mechanism and the final states in outgoing nuclei. The high-lying resonances in $^{14}$C are reconstructed for three sets of well discriminated final states in $^{10}$Be. The results confirm the previous decay measurements with clearly improved decay-channel selections and show also a new state at 23.5(1) MeV. The resonant states at 22.4(3) and 24.0(3) MeV decay primarily into the typical molecular states at about 6 MeV in $^{10}$Be, indicating a well developed cluster structure in these high-lying states in $^{14}$C. Further measurements of more states of this kind are suggested.
\end{abstract}



\footnotetext[0]{\hspace*{-3mm}\raisebox{0.3ex}{$\scriptstyle\copyright$}2013
Chinese Physical Society and the Institute of High Energy Physics
of the Chinese Academy of Sciences and the Institute
of Modern Physics of the Chinese Academy of Sciences and IOP Publishing Ltd}%

\begin{multicols}{2}

\section{Introduction}
It has been well established that clustering is one of the fundamental aspects in structuring the light nuclei\cite{Oer2006,Ye2012},
although it appears also quite frequently in heavy nuclei, such as the $\alpha$ particle formation and decay\cite{Ren2013}.
For isospin-asymmetrical nuclei, the presence of valence nucleons surrounding the core clusters may help to stabilize the system, similar to the molecular structure for atomic systems\cite{Oer2006, Ye2012, Hor2012}. It is speculated that neutron-excess beryllium isotopes are good candidates for molecular
structure due to its richness of valence neutron configurations built on a well established $\alpha + \alpha$ rotor of
$^8$Be\cite{Yan2014, Yan2014-b, Yan2015}. A natural extension to such ideas goes to carbon isotopes, which may have three centers
located by three tightly-bound $\alpha$-particles\cite{Oer2006}. Theoretically both triangle and linear-chain configurations were proposed
for carbon isotopes, based on the molecular-type potential between two $\alpha$ clusters\cite{Oer2006,Wik1986}.

The famous cluster state at an
excitation energy of 7.65 MeV in $^{12}$C, the so called Hoyle state, was identified as an $\alpha$-particle Bose-Einstein (BES)
condensation where all $\alpha$-particles stay at the relative S-state\cite{Oer2006}. It is predicted that the addition of valence
neutrons into the $\alpha$-conjugate nuclei may help to stabilize the chain-state. Expectations were given to $^{14}$C and $^{16}$C
isotopes based on the state-of-art model calculations\cite{Oer2006, Tad2010}. Experimentally the measurements on $^{14}$C have been
carried-out quite intensively during the past decade. In 2003, Soic et al measured the excited states in $^{14}$C up to 24 MeV via the
cluster transfer reaction of $^7$Li(${^9}$Be, $^{14}$C)$^2$H \cite{Soi2003}. The decay of the high-lying excited states in $^{14}$C
into the $^{10}$Be + $\alpha$ final channel were distinguished according to the reaction $Q$-values corresponding to the ground- and
first excited-states (3.4 MeV, $2^+$), and the states close to 6 MeV, in $^{10}$Be. The four excited states at around 6 MeV
($2^+, 1^-, 0^+, 2^-$) in $^{10}$Be possess typical molecular structure built on 2-$\alpha$ cores, with both
$\sigma$- and $\pi$-bindings \cite{Oer1997}. Thus based on the structural link between the mother and daughter nuclei, the states at 22.4 and 24.0 MeV
in $^{14}$C were considered to have a structure change compare to other states \cite{Soi2003}, since they decay preferably into the 6 MeV states in $^{10}$Be. The drawback of this earlier measurement is the large background below the Q-value peaks, which prohibited the reconstruction of the mother nucleus from a clear identification of the decay process. Since then a number of 2n transfer or 2p removal experiments were also performed to populate the high-lying excited states in $^{14}$C, but they are basically sensitive to the single-particle configurations with oblate or prolate deformations\cite{Mil2004,Oer2004,Pri2007,Hai2008}. The latest resonant $\alpha$-scattering experiments are dedicated to search for cluster states in $^{14}$C. Based on the observed resonant states some possible molecular-rotation bands or linear-chain states were proposed \cite{Fee2014,Fri2016}. However
due to the limited energy resolution of this thick-target method the decays to 6 MeV states in $^{10}$Be were not separated.

In the present work, we have selected the reaction $^9$Be($^9$Be, $^{14}$C)$^4$He, which has an extremely large $Q$-value compared to other competing reaction channels. The high-lying states in $^{14}$C are reconstructed from the $^{10}\rm{Be} + \alpha$ decay channels respective to the ground-, 2$^+$ and $\sim$ 6 MeV states in $^{10}$Be.

\section{Experiment}

\begin{center}
\includegraphics[width=8 cm]{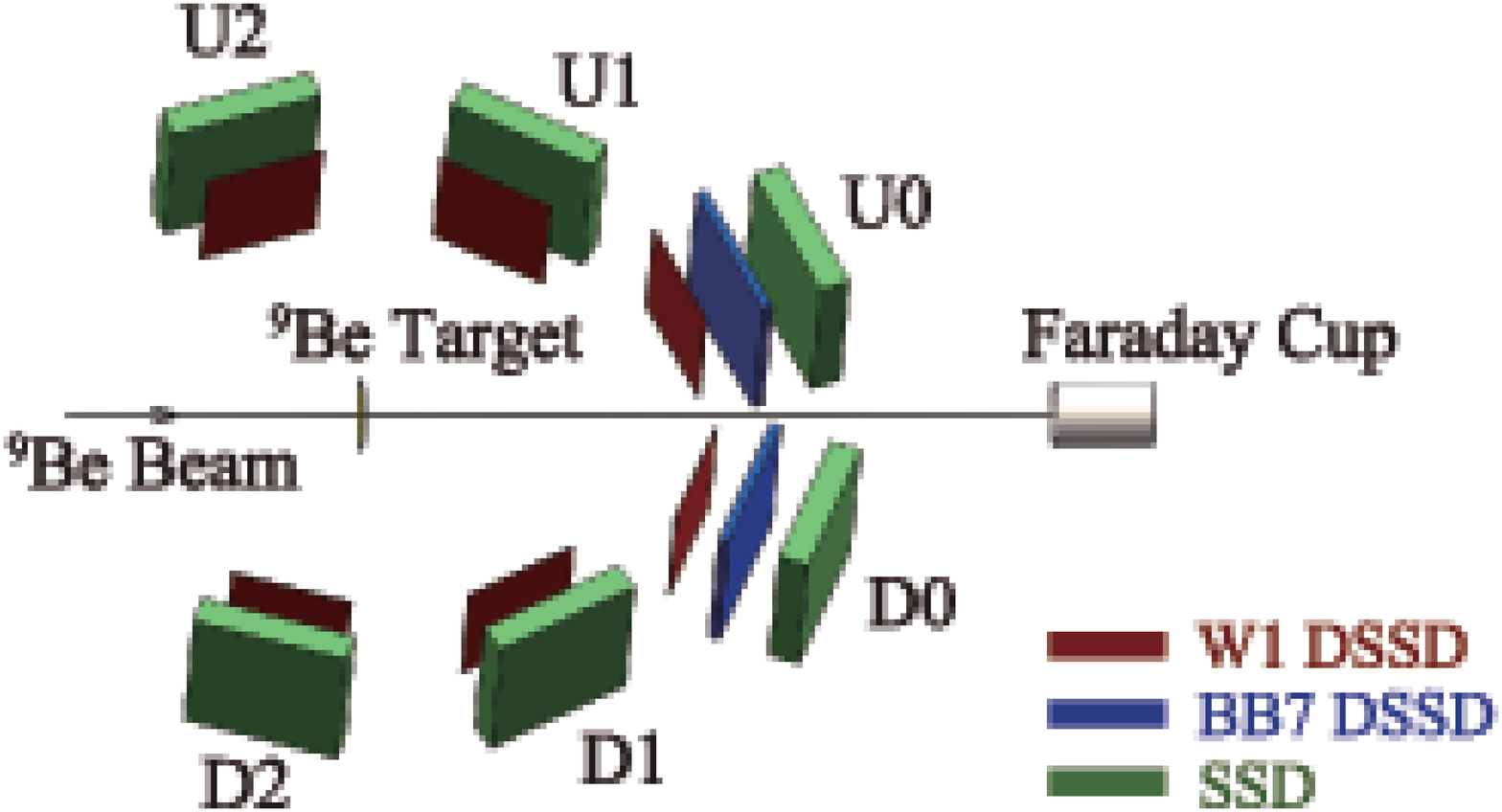}
\figcaption{\label{fig1} A schematic view of the experimental setup.}
\end{center}

The experiment was performed at the HI-13 tandem accelerator at China Institute of Atomic Energy (CIAE). A $^9$Be beam at 45 MeV and with an intensity of about 7 enA was incident on a 0.9 $\rm{\mu}$m thick self-supporting $^9$Be target. A schematic diagram of the detection system is shown in Fig.~\ref{fig1}. Reaction products were detected by six particle telescopes, namely U0, U1, U2, D0, D1, D2. The front faces of the forward telescopes (U0 and D0) were located at 140 mm from the target and centered at $\pm$23¡ã relative to the beam direction. Each of them consisted of one double-sided silicon strip detector (DSSD) with a thickness of 60 $\rm{\mu}$m (W1-60 ), one DSSD with a thickness of 500 $\rm{\mu}$m (BB7-500) and one large-size silicon detector (SSD) with a thickness of 1500 $\rm{\mu}$m (MSX40-1500). The sensitive area is 50 mm $\times$ 50 mm for W1-60, and 64 mm $\times$ 64 mm for BB7-500 and MSX40-1500. Both front and back faces are divided into 16 or 32 strips for W1-60 or BB7-500, respectively. Each of the other four telescopes consisted of one W1-60 and one MSX40-1500. U1 and D1 were located at a distance of 116 mm from the target, and centered at angles of $\pm$60$^0$, respectively, while U2 and D2 at 114 mm and $\pm$109$^0$. The forward telescopes (U0 and D0) were primarily used to detect the decay fragments from the excited $^{14}$C, while others for the detection of the recoiled $\alpha$-particles. Energy calibration of the detectors was realized using a combination of $\alpha$ sources. The energy match for all silicon strips in one detector was achieved according to the uniform calibration method described in \cite{Qia2014}.

\section{Analysis and results}

The particle identification (PID) up to carbon isotopes were achieved by the applied telescopes thanks to the excellent energy resolutions of the silicon detectors, which were typically less than 1.0$\%$ for 5.49 MeV ¦Á particles from a $^{241}$Am source. We then selected events with $^{10}$Be + $\alpha$ two-fold coincidence, taken from the two opposite forward telescopes (U0 and D0). The remaining undetected nucleus is expected to be an $\alpha$-particle, of which the momentum and energy can be deduced according to the respective conservation requirements. The reaction $Q$-value can then be calculated using the standard definition:
\begin{eqnarray}
\label{eq1}
Q &=& E_{\rm{tot}} - E_{\rm{beam}}\nonumber \\
&=& E_{\rm{10Be}} + E_{\alpha 1} + E_{\alpha 2} - E_{\rm{beam}}
\end{eqnarray}
As shown in Fig.~\ref{fig2}, the $Q$-value for final particles all at their ground states, $Q_{\rm{ggg}}$, is 5.24 MeV for the present work (Fig.~\ref{fig2}(b)) and -1.91 MeV for the previous measurement\cite{Soi2003}. An extremely large $Q$-value, due to two deeply-bound $\alpha$-particles in the final channel, is characteristic in our experiment, and helps to clean up the reaction mechanism. In the figure, the peak next to $Q_{ggg}$ corresponds to the first excited state (3.4 MeV, 2$^+$), whereas the further next one at about 6 MeV excitation means the population of the four adjacent states, in $^{10}$Be \cite{Soi2003}. Since other possible reaction channels together with corresponding excitations lead to much lower $Q$-values, the peaks in Fig.~\ref{fig2}(b) have almost no contaminations. The broad background spectrum below the peaks is much less significant in the present case (Fig.~\ref{fig2}(b)) than in the previous work (Fig.~\ref{fig2}(a)). The shape of the present $Q$-value spectrum was verified by using the completely detected 3-fold coincident events. It is evident that the present measurement provides an opportunity to unambiguously identify the resonant states in $^{14}$C according to its decay to each final state in $^{10}$Be.
\begin{center}
\includegraphics[width= 9 cm]{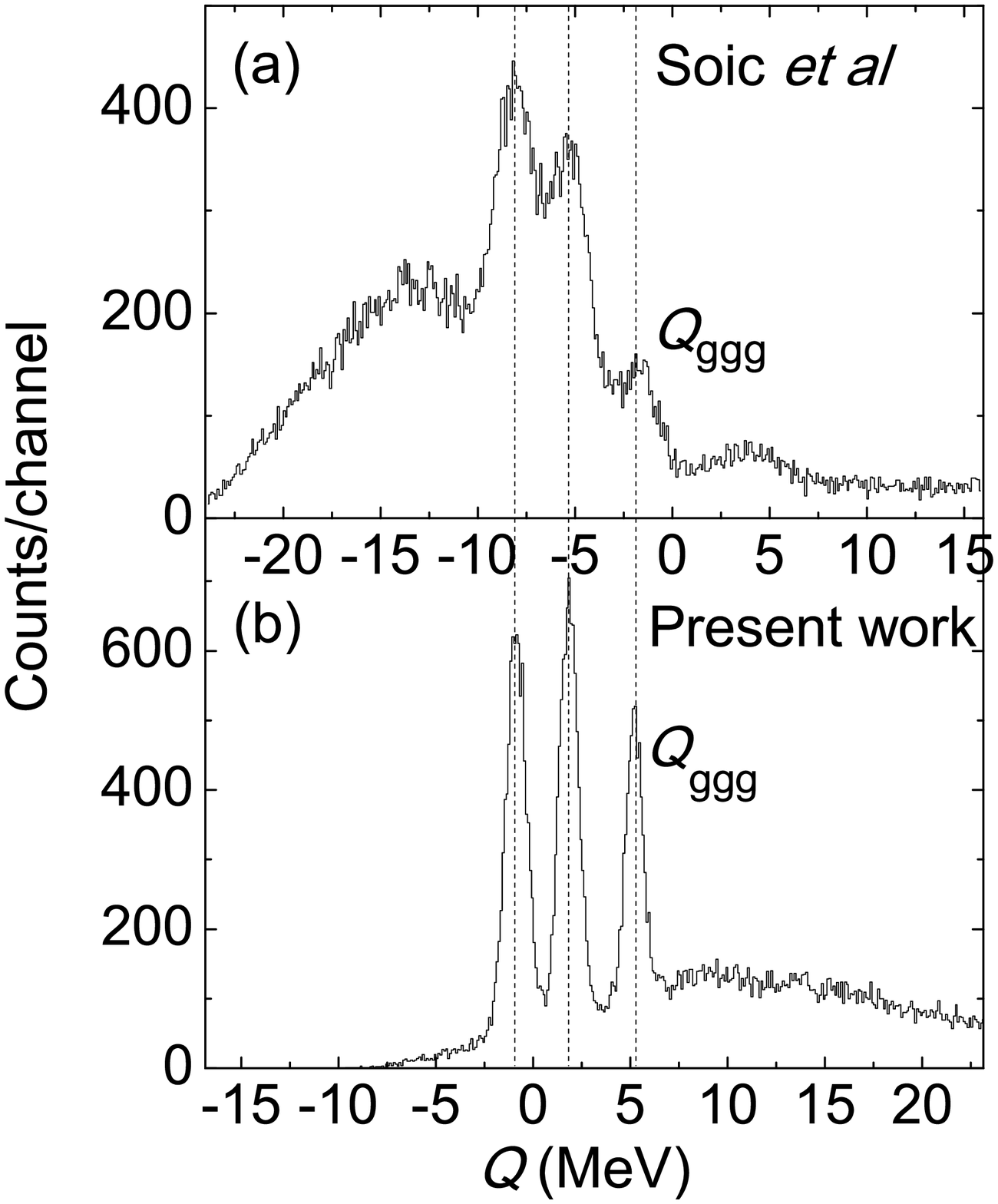}
\figcaption{\label{fig2} $Q$-value spectra for (a) the reaction $^7\rm{Li}(^9\rm{Be},\alpha^{10}\rm{Be})^2\rm{H}$ \cite{Soi2003} and (b)
the present measurement of $^9\rm{Be}(^9\rm{Be},\alpha^{10}\rm{Be})\alpha$. The spectra are aligned according to the corresponding peaks.
The thin vertical dashed lines are used to guide the eyes.}
\end{center}
Gated on the three $Q$-value peaks shown in Fig.~\ref{fig2}(b) and using the method described in \cite{Yan2015, Lv2011}, the high-lying resonant states in $^{14}$C are reconstructed from the $^{14}\rm{C} \rightarrow \alpha + ^{10}\rm{Be}$ cluster-decay channel, and the results are presented in Fig.~\ref{fig3}(b), together with previous results (Fig.~\ref{fig3}(a)) for comparison. The results are plotted as empty diamonds, filled circles and empty triangles for decaying to the ground, first excited and $\sim 6$ MeV states in $^{10}$Be, respectively. For all three cases it can be seen that our detection acceptance (efficiency) is in favor of the higher energy side, resulted from the angular coverage of our detector setup.
\end{multicols}
\ruleup
\begin{center}
\includegraphics[width= 12 cm]{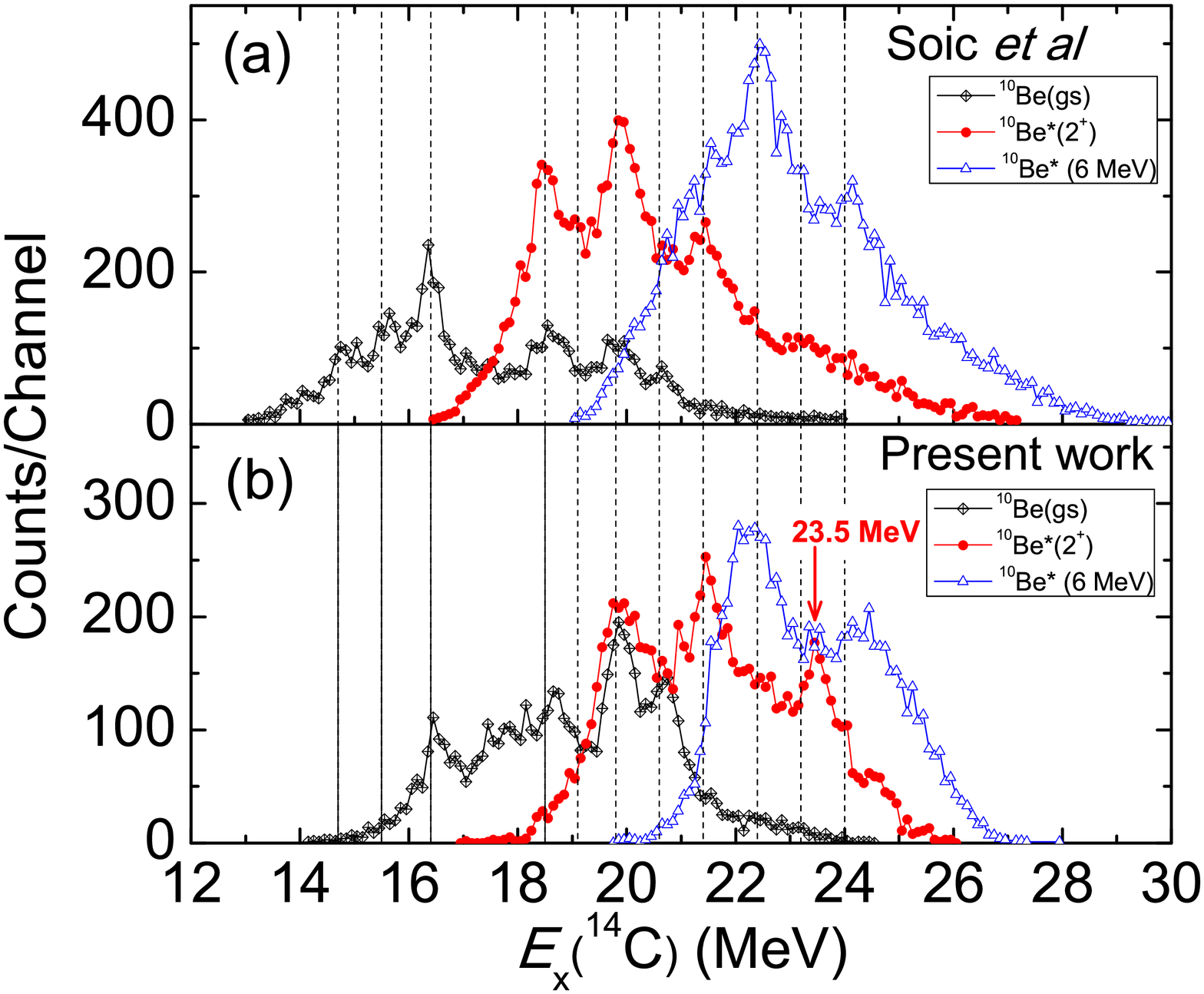}
\figcaption{\label{fig3} The excited states in $^{14}\rm{C}$ reconstructed from the $\alpha + ^{10}\rm{Be}$ cluster-decay channel, with
the upper panel (a) taken from the previous measurement\cite{Soi2003} and the lower panel (b) from the present work. The decay final states
are selected for the ground, first excited and $\sim 6$ MeV states in $^{10}$Be.}
\end{center}
\ruledown
\begin{multicols}{2}
From Fig.~\ref{fig3} it can be seen that most states coincide very well with each other for the two measurements, demonstrating the reliability of the present experimental and data analysis methods. In case of decaying to $^{10}\rm{Be}_{gs}$ (empty diamonds), the previously observed strong population of the 16.4 MeV state in $^{14}$C is well confirmed in our experiment, while the peaks at 19.8 MeV and 20.6 MeV are much more significant in the present measurement. As for $^{10}\rm{Be}(3.4 \rm{MeV}, 2^+)$ final state (filled circles), in addition to the confirmation of states at 19.8 and 21.4 MeV, we have clearly observed a new peak at 23.5(1) MeV, as indicated by the arrow, which appeared very weakly in the previous measurement due to the low acceptance there. The uncertainty of the peak positions for our experiment is about 80 keV, dominated by the systematic errors in the energy calibration of the telescopes (about 50 keV in the interested energy range) and in the estimation of the dead-layer thickness of the silicon detector ( $1.0 \pm 0.3 \rm{\mu m}$ at each surface). Again we see a good match of the $^{14}$C-states which decay simultaneously into $^{10}\rm{Be}_{gs}$ and $^{10}\rm{Be}(3.4 \rm{MeV}, 2^+$). Most of these resonant states have also been observed in recent resonant scattering experiment \cite{Fee2014,Fri2016}, where spin assignment were made using the angular correlation method. The states at very high energy of 23.5(1) MeV has only been identified in the present experiment, which might be an extension of the tentatively proposed molecular rotation band (see Fig.12 in Ref.\cite{Fee2014}).

 The most high-lying states at 22.4(3) and 24.0(3) MeV (empty triangles in the Fig.~\ref{fig3}) decay primarily into the $\sim 6$ MeV states in $^{10}$Be, although a small kink appears also in the spectrum for decaying to $^{10}\rm{Be}(2^+)$, implying a possible configuration mixing. It should be noted that the uncertainty of the peak positions here is much larger due to the energy deviation of the four states around 6 MeV in $^{10}$Be. It is widely accepted that these $\sim$ 6 MeV states in $^{10}$Be are band heads of reflection-asymmetric molecular-rotation \cite{Oer1997}. Considering the structural-link in the decay, it is reasonable to consider these very high-lying resonances in $^{14}$C as well developed three-center molecular states which overlap strongly with the specific structure of its subcomponent $^{10}$Be. We confirmed two states of this kind here with much clearer $Q$-value selection as demonstrated in Fig.~\ref{fig2}. It is expected that more states of this kind might be observed at lower as well as higher energy sides of the present spectrum (triangles in Fig.~\ref{fig3}) if we extend our measurement to smaller and larger angles, respectively. It is also important to determine the spins of these states based on the experimental information.

\section{Conclusion}
Using entirely silicon detectors with excellent energy resolution and particle identification performance, we are able to reconstruct the the high-lying resonant states in $^{14}$C from the $^{10}\rm{Be} + \alpha$ cluster-decay channel. The applied reaction, namely $^9\rm{Be}(^9\rm{Be},\alpha^{10}\rm{Be})\alpha$, is of extremely large $Q$-value, making it an excellent case to select the reaction mechanism and the final states in the outgoing nucleus. The contamination and background are largely reduced compared to previous measurements. The high-lying excited states in $^{14}$C are reconstructed, corresponding to three sets of well discriminated final states in $^{10}$Be. Most of the presently observed states, which decay to $^{10}\rm{Be}_{gs}$ and to $^{10}\rm{Be}(3.4 \rm{MeV}, 2^+)$, agree very well with the previous observations \cite{Soi2003}, while a state at 23.5 MeV is clearly identified for the first time. The states at 22.4 and 24.0 MeV decay primarily into 6 MeV states in $^{10}$Be. The latter possess typical molecular structure, implying a well developed three-center molecular configuration in the mother states, considering the structural link in the decay. The present results, based on the significantly improved selection of reaction and decay mechanisms, greatly support the richness of cluster formation and decay in excited states of light neutron-rich nuclei. More measurements of this kind, especially with the spin determination, are certainly encouraged.

\acknowledgments{The authors would like to thank the staffs at the HI-13 tandem accelerator for their excellent work in providing the beams. This work is supported by the 973 Program of China (No. 2013CB834402) and the National Natural Science Foundation of China (No. 11275011, 11535004).}

\end{multicols}

\vspace{-1mm}
\centerline{\rule{80mm}{0.1pt}}
\vspace{2mm}

\begin{multicols}{2}

\end{multicols}

\clearpage
\end{CJK*}
\end{document}